\documentclass[conference]{IEEEtran}

\usepackage{url}
\usepackage{graphicx}
\usepackage{multirow}   
\usepackage{enumitem}
\usepackage{xspace}  
\usepackage{verbatim}   
\usepackage{algpseudocode}
\usepackage{lipsum}
\usepackage{float}
\usepackage{graphicx}
\usepackage{flushend}
\usepackage{caption}
\usepackage{changepage}
\usepackage{scrextend}
\usepackage[square,numbers]{natbib}

\begin{document}
\title{Comment Generation for Source Code: Survey}

\author{\IEEEauthorblockN{Xiaoran Wang and Benwen Zhang}
\IEEEauthorblockA{
xiaoranlr@gmail.com, benwen@udel.edu}
}

\maketitle

\begin{abstract}
Research has shown that most effort of today's software development is maintenance and evolution. Developers often use integrated development environments, debuggers, and tools for code search, testing, and program understanding to reduce the tedious tasks. One way to make software development more efficient is to make the program more readable. There have been many approaches proposed and developed for this purpose. Among these approaches, comment generation for source code is gaining more and more attention and has become a popular research area. In this paper, the state of art in comment generation research area are summarized and the challenges and future opportunities are discussed.

\end{abstract}


\section{Introduction}

Research has shown that more than 60\% of software engineering resources are spent on maintenance~\cite{erlikh00}. Software maintenance is the process of modifying a software system after delivery to fix bugs, improve performance, or adapt to a changing environment~\cite{lientz78}. Software maintenance requires code comprehension~\cite{981648}, as reading and understanding source code is the prerequisites of any modification. Program comprehension is time-consuming and cost most of developers' time~\cite{goldberg-1987,murphy-ecoop-05,Haiduc,Deimel:1985,Raymond,readingcode3}.  Developers often use integrated development environments, debuggers, and tools for code search, testing, and program understanding to reduce the tedious tasks. 
  
To reduce the effort of program comprehension and software maintenance, many tools have been developed to support the tedious and error-prone tasks~\cite{prog-comp-survey-05, fifteen-years-of-pc}. Formatting source code naturally helps program comprehension and there has been some research toward this direction~\cite{wang2011automatic,wang2013automatic,Jackson200840,wang-thesis}. More prevalently, researches have been conducted to generate comments to help program comprehension~\cite{sridhara2010towards,giri11,wang-2017-orau}. Comment generation for source code is gaining more and more attention and has become a popular research area.
In this survey, the state of art in the comment generation research area is summarized.

\section{State of the Art}

The researches of comment generation for source code can be categorized to three main classes based on the main technique utilized: 1) informaiton retrieval, 2) program structure information and 3) software artifacts beyond source code. This section first describes the work under the three categories and then summarize the fundamental natural langauge processing research in software engineering.

\subsection{Information Retrieval}

Researchers have integrated natural language processing techniques to analyze the natural language clues in the source code in recent years~\cite{lawrie2011expanding,
enslen2009mining,madani2010recognizing}. 
Information retrieval(IR) techniques use words in documents to determine the similarity between documents and queries. IR calculates a similarity score between a query and documents and the results of a query according to relevance are ranked based on the scores~\cite{shepherd07,wang2008categorization}. 

IR has become a standard technique to address the problem in natural language domain. Since its great promise in natural language, many researchers have applied IR to locate concepts~\cite{Rajlich:2004,zhao2006sniafl,haiduc2013automatic,hill2009automatically,marcus2004information,petrenko2013concept,peng2013improving,wang2013improving} and reconstruct documentation traceability links in source code~\cite{marcus2003recovering,antoniol2002recovering,Maskeri2008,Ohba2005}. These work use different variations of IR techniques to link the query with source code or build links between different software artifacts. The concepts located and links built can be served as the comment for the linked source code. 
IR-based concern location tools treat source code as a ``bag of words'' and focus on the individual words in the code. Research that uses IR does not generate natural langauge phrases or sentences. Instead, the generated comments are a bag of words. 

There have been many research towards extracting topic words or verb phrases from source code~\cite{Maskeri2008,Ohba2005,hill2009automatically}. They want to identify code fragments that are related to a given action or topic. Other research tries to cluster program elements that share similar phrases~\cite{kuhn07}.  These methods rely solely on the linguistic information to determine the topic of the code segment, which is often not adequate for describing the source code. 

Because IR approaches treat source code as a bag of words, the structure information is missing. More importantly, towards this research direction, the comment generated are not real ``comment''. At least, not like the comment developers put in the source code. 

\subsection{Program Structure Information}

In 2010, Sridhara et~al.~\cite{sridhara2010towards} first introduce an approach to generate summary comments for Java methods using program structure information. The basic idea is selecting the important statements from a method and then translate the statements into natural language phrases. They first build data flow graph of the method and then analyze the data flow to identify the important statements.

Later, Sridhara et~al.~\cite{sridhara2011automatically} introduced high level action in 2011. A high level action is defined as a high level step of what the code does. They identify a known set of high level actions based on a set of predefined templates that frequently occur in the source code. While the approach identifies code fragments that implement high level algorithmic steps, 
 the technique is limited in the kinds of high level action it is able to identify. Only 24\% of switch blocks, 40\% of if-else blocks, and  15\% of iterator loops implemented one of the templates. Later, Wang et al.~\cite{wang2011automatic} propose an automatic approach to cover loops without manually developing rules. They focus on identifying high level actions that are implemented by loop structures. While the work of Wang et al. advances the techniques, their technique is limited to loops only. After that, the same authors developed an approach to cover the more prevalent code fragments in source code - data flow chain~\cite{wang-orau}.
 
Complementary to the work of Sridhara et~al.~\cite{sridhara2010towards}, the same authors also presented a technique to generate comments for parameters and integrated those descriptions with method summaries~\cite{5970165}. Beyond method level, Moreno et al. developed an approach that generates summaries for Java classes~\cite{6613830}.

\subsection{Software Artifacts Beyond Source Code} 
The work that falls into ``program structure information'' category only uses the source code as the learning source to develop rules or build models. There have been some researches using forums to get the natural language clues, as developers often discuss the problem online in the site like Stack Overflow~\cite{7884638,Chatterjee:2017}..

Wong et al. used question and answer sites for automatic comment generation~\cite{6693113}. They extracted code-description pairs from the question title and text and used code clone detection to find source code snippets that are almost identical to the code of the code-description mappings. The code fragment in the pairs are often code copied from developers' projects and posted by developers. The basic idea is identifying comments for code fragments and reusing the comments for other projects. However, this technique only works for a small fraction of code fragments. The approach is scalable to millions of projects, but would not generate many comments for an individual project.

Ying and Robillard presented a supervised machine learning approach that classifies whether a line in a code fragment should be in a method summary~\cite{Ying:2013}. However, the generated summary is not the natural language but a set of statements.

\subsection{Fundamental Natural Language Processing Techniques in Comment Generation}

Software engineering tools often use part-of-speech (POS) taggers to identify the POS of a word and tag the word as a noun, verb, preposition, etc. and then chunk the tagged words into grammatical phrases to help distinguish the semantics of the component words.
Traditional taggers trained on natural language texts work well on the news and natural language artifacts, but their accuracy on source code reduces as the input moves farther away from the highly structured sentences found in traditional newswire articles.

There have been researches towards building appropriate POS taggers for software engineering researches. Binkley et al.~\cite{binkley2011improving} presented a POS tagger for field names in source code. They produce sentences with templates. Falleri et al.~\cite{falleri2010automatic} use the TreeTagger~\cite{schmid2013probabilistic}, a tagger trained on English text, to perform the POS tagging. Sridhara et al.~\cite{giri08} use well known semantic similarity techniques and perform well on English text~\cite{budanitsky2006evaluating,martin2000speech,pedersen2005maximizing}. 

Many software related words are not in the source code itself, but in the various associated software artifacts, such as the online forum, bug reports, commit logs, email communications, etc. Tian et al.~\cite{tian2014automated} developed an automatic approach that builds a software specific WordNet like database by leveraging the contents of posts in Stack Overflow. They measure the similarity of words by computing the similarities of the weighted co-occurrences of these words in the textual corpus. In addition to the work of mining semantically-similar Words, Wang et al.~\cite{wu2009learning} infer semantically related tags from FreeCode. Falleri et al.~\cite{falleri2010automatic} showed how to extract a network of identifiers connected by “is-more-general-than” or “is-a-part-of” relationships from source code. 
 
Some researchers have developed automatic techniques to capture co-occurring word pairs~\cite{maarek1991information,manning1999foundations}, but co-occurrences do not capture the information about the nature of the relationship between words beyond that the words co-occur in the same context. 
However, others have leveraged the POS tagging to analyze identifiers and build models of the usage of words in identifiers. 
Shepherd et al. showed that representing method identifiers as the verb and direct object pairs can improve search by focusing on the actions and what they action upon~\cite{shepherd07}.

\section{Challenges and Opportunities}

One of the drawbacks of the current approach is that they only generate descriptive comments. These comments are commonly used but do not reflect the knowledge beyond the source code. The comment that provides developers design intuition and thoughts are equally meaningful as description comments. There has been work about \texttt{TODO} comments, but such kinds of work are very limited. In addition, different coders may need different kinds of comments to understand the source code. Intelligent and customized comments are in high demand.


In the current state, developers often use their own dataset as their subjects in the studies. The techniques using a different dataset is often difficult to compare with another technique. Moreover, many research uses human evaluation to judge the quality of the results. This increases the difficulty to build a standard benchmark.


Previous work on comment generation in the high level action approach often use heuristics to come up with rules. One reason is the lack of high-quality training data. As common sense, developers write comments less frequently than expected. Building a high-quality dataset would not only help researchers develop an automated approach based on machine learning, but also lead to more standard matrics for evaluation and comparison.


Recurrent Neural Networks (RNNs) are gaining a lot of attention in recent years because it has shown great promise in many natural language processing tasks. Despite their popularity, to my knowledge, there has not been work using RNNs to generate comments. The nature of RNNs advanced natural language generation and translation. It seems fit in comment generation also, as descriptive comments are essentially summarization and translation of source code to natural language. 


As labeled data is always difficult to obtain. High-quality data limit the application of supervised learning methods. In machine learning, there have been many unsupervised methods that might be explored on source code data. These methods include clustering, latent semantic analysis, latent Dirichlet allocation, etc. These methods could potentially provide labels for the supervised learning methods.

Besides the progress made by many types of research in the area of comment generation, the state of the art techniques are still very limited.  This survey serves only a start of exploration of this area. 

\bibliographystyle{plain}
\bibliography{main} 

\end{document}